\documentclass[useAMS,usenatbib,usegraphicx]{mn2e}

\newcommand{\cm}{cm$^{-1}$}
\newcommand{\teff}{T$_{\rm eff}$}
\newcommand{\mim}{$\mu$m}

\newcommand{\HL}{H\"{o}nl--London}

\newcommand{\JMS}{J. Mol. Spect.\ }
\newcommand{\JCP}{J. Chem. Phys.\ }

\newcommand{\AAA}{A\&A}
\newcommand{\AAAS}{A\&AS}
\newcommand{\ApJ}{ApJ}
\newcommand{\AJ}{Astron. J}

\newcommand{\MNRAS}{MNRAS}
\newcommand{\leen}{lab/empirical energy level list}
\newcommand{\lee}{lab/empirical}
\newcommand{\ai}{{\em ab initio}}
\newcommand{\wzcas}{WZ~Cas}
\newcommand{\mum}{$\mu$m}
\newcommand{\dpr}{$^{\prime\prime}$}
\newcommand{\pr}{$^{\prime}$}
\newcommand{\lH}{$l/H$~}

\title[Improved linelist for HCN/HNC]{Improved HCN/HNC linelist, model atmospheres and synthetic spectra for \wzcas.}

\author[Harris et al.]{G. J. Harris$^1$, J. Tennyson$^1$, B. M. Kaminsky$^2$, Ya. V. Pavlenko$^{2,3}$, and H. R. A. Jones$^3$\\
$^1$ Department of Physics and Astronomy, University College London, London, WC1E 6BT, UK.\\
$^2$ Main Astronomical Observatory, National Academy of Sciences, Zabolotnoho 27, Kyiv-127 03680, Ukraine.\\
$^3$ Centre for Astrophysics Research, University of Hertfordshire, Hatfield. AL10 9AB, UK. }

\begin{document} 

\maketitle
                                         
\begin{abstract}

We build an accurate database of 5200 HCN and HNC rotation-vibration energy levels, determined from existing laboratory data. 20~000 energy levels in the \citet{linepaper} linelist are assigned approximate quantum numbers. These assignments, lab determined energy levels and \citet{linepaper} energy levels are incorporated in to a new energy level list. A new linelist is presented, in which frequencies are computed using the lab determined energy levels where available, and the \ai\  energy levels otherwise.

The new linelist is then used to compute new model atmospheres and synthetic spectra for the carbon star WZ~Cas. This results in better fit to the spectrum of WZ~Cas in which the absorption feature at 3.56 \mum\  is reproduced to a higher degree of accuracy than has previously been possible. We improve the reproduction of HCN absorption features by reducing the abundance of Si to [Si/H] = --0.5 dex, however, the strengths of the $\Delta v=2$ CS band heads are over-predicted.

\end{abstract}


\section{Introduction.}
\label{sec:Intro}

Cool AGB carbon stars often show a strong absorption feature at 3 microns. This feature was discovered \citep{Johnson} and later identified as the C-H stretch mode of HCN and/or C$_2$H$_2$ \citep*{Fay,Ridgeway}. Subsequent to the discovery of HCN in observed carbon star spectra, it was found that the line opacity of HCN has a strong effect upon the structure of model atmospheres \citep{Eriksson,Jorgensen1}. \citet*{linepaper}, referred to as HPT, have made publicly available an extensive and accurate {\em ab initio} HCN and HNC linelist. This linelist has been used to compute model atmospheres and synthetic spectra that reproduce the HCN features in the observed spectra of the carbon stars TX~Psc and WZ~Cas and has also allowed the identification off HNC absorption at 2.9 \mum\  \citep{paper1}. However, the HPT linelist has inherent inaccuracies which result in the line frequencies deviating from laboratory measurements by 3 \cm\  or more. This error is sufficiently large to be noticeable at the resolving power of the ISO SWS spectrometer. In particular \citet{paper1} fail to accurately reproduce the observed HCN Q-branch absorption feature at 3.56-3.62 \mum.

In this work we assign approximate quantum numbers to 20~000 of the HPT \ai\  energy levels. The quality of the HPT linelist is improved by substituting energy levels derived from laboratory measurements of line frequencies, for \ai\  energy levels. This improves the frequencies of the lines for which experimental data is available, to spectroscopic precision (0.1 \cm) and allows the calculation of more accurate C-star synthetic spectra. These new synthetic spectra significantly improve the reproduction of the HCN Q-branch absorption feature at 3.56-3.62 \mum.

\section{Energy level assignments}
\label{sec:assign}

\begin{figure*}
\includegraphics[angle=0,width=84mm]{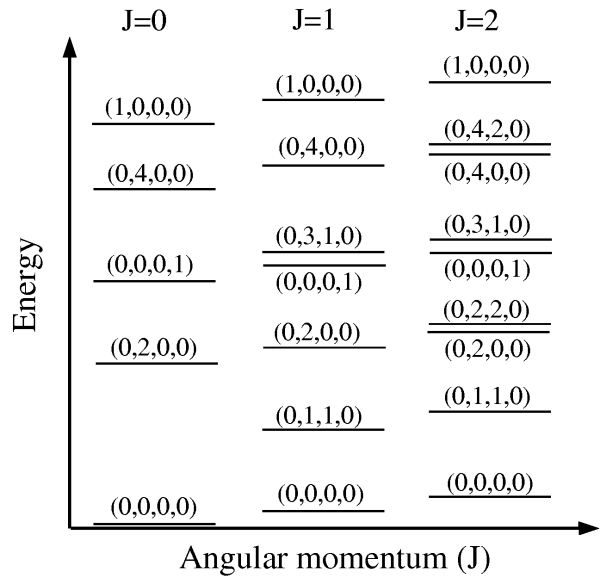}
\caption{A schematic showing the distribution of even parity HCN energy levels as a function of angular momentum (J) and Energy. States are labelled ($v_1$,$v_2$,$l$,$v_3$).}
\label{fig:levs}
\end{figure*}

The linelist of \citet{linepaper} (HPT) is calculated using the only exact quantum numbers for a hetronuclear triatomic, angular momentum ($J$) and parity. Conversely laboratory data is generally presented in terms of the approximate vibrational quantum numbers C-H stretch ($v_1$), bend ($v_2$), C-N stretch ($v_3$), vibrational angular momentum ($l$) and isomer (HCN or HNC). Where $l$ can take values from 0 or 1, if $v_2$ is even or odd, and up to  $l\le v_2$ and $l\le J$, in steps of 2. Thus for a $v_2=4$ with $J\ge4$, $l$ can take values of 0,2,4, but for $J=2$ or $J=3$, $l$ can only take values of 0 or 2.
Figure \ref{fig:levs}, is a schematic showing the distribution of even parity energy levels as a function of J and energy. New energy levels appear with each higher $J$, these are the $l=J$ states. In order to match the lab determined energy levels with their corresponding \ai\  energy levels it is necessary to assign the approximate quantum numbers to the \ai\  energy level list.

By comparison with the assignments of \citet{Bowman} the non-rotating ($J=0$) vibrational states were assigned by hand up to 10~000 \cm\  above the zero point energy. We then least squares fitted the vibrational expansion:
\begin{eqnarray}
E(v,l)+E_0 & = & \sum^3_{i=1} \omega_i(v_i+d_i/2) \nonumber \\
           & + & \sum^3_{i=1}\sum^3_{j=1} x_{ij}(v_i+d_i/2)(v_j+d_j/2) \nonumber \\
           & + & x_ll^2
\label{eq:vibexp}
\end{eqnarray}
to the assigned energy levels. Where $E_0$ is the zero point energy, $\omega_i$ are the harmonic constants, $x_{ij}$ are anharmonic constants, the degeneracies are $d_1=d_3=1$ and $d_2=2$. HCN and HNC states were fitted separately. For $J=0$, $x_l$ is zero.
As found by \citet{Maki01} these vibrational expansions give poor fits to both HCN and HNC, however the fits were useful in identifying and correcting mis-assigned states. 

 A somewhat different assignment strategy is required, for $J>0$, as the $l>0$ states do not occur for $J=0$. The distribution of energy levels in a vibrational band of HCN can be approximated by using the rotational expansion for a linear triatomic molecule:
\begin{eqnarray}
E(v,J) & = & E(v)+B_v[J(J+1)+l^2] \nonumber \\
       &   & -D_v[J(J+1)+l^2]^2  \nonumber \\
       &   & +H_v[J(J+1))+l^2]^3 + ... 
\label{eq:rotexp}
\end{eqnarray}
where E(v) is the vibrational energy of the band, $B_v$ is the rotational constant, $D_v$ is the centrifugal distortion constant and $H_v$ is a higher order constant.
To estimate the $J=1$, $l=0$ energy levels, we used the ground vibrational state values of the rotational constant ($B_v$) of 1.478221 and 1.512112 \cm\  for HCN and HNC respectively, and the first two terms of eq. \ref{eq:rotexp}.
By comparing these estimated energy levels with the \ai\  levels the $J=1$, $l=0$ states were assigned. The remaining states were those with $l=1$, the lowest energy of these states is the (0,1$^1$,0), bending fundamental which was assigned.
The vibrational expansion eq. \ref{eq:vibexp} was used to fit the newly assigned $J=1$, $l=0$ and (0,1$^1$,0) states and thus estimate values for the higher $J=1$, $l=1$ states. Again, by comparison of the estimated values with the \ai\  values, the $J=1$, $l=1$ states were assigned.

Higher J states were assigned in the same way, the rotational expansion was fit to the previously assigned data for each band, the $l<J$ states assigned by comparison. These assigned $l<J$ states were fitted with the vibrational expansion, allowing the assignment of the $l=J$ states. This process was carried out on a $J$ by $J$ basis until energy levels up to $J=60$ had been assigned. In total we assigned approximate quantum numbers to around 20~000 energy levels, for HCN this is up to 3 quanta of H-C stretch, 13 quanta of bend and 4 quanta of C-N stretch. 

The distribution of the line strengths of a given band can be approximated by \HL\  factors. By fitting the line transition dipoles of each band using  \HL\  factors is is possible to identify lines which have an intensity which is inconsistent with that of the band as a whole. In this way we cross checked the consistency of our assignments, using line intensities, and corrected our assignments where necessary. However, there are states in the linelist of HPT which are close enough in both energy and quantum numbers for intensity stealing to break the distribution of intensities in a band, see \citet*{spectpaper}. For these states the assignments of approximate quantum numbers, based on energy and/or transition dipole is difficult and arguably meaningless, such states are left unassigned. 

\section{Energy levels determined from laboratory data.}
\label{sec:expen}

HCN and HNC have been studied in detail in the laboratory \citep{Smith,Maki00,Lecoutre,Northrup,Maki01}. HCN has been studied up to a maximum of 6 quanta in the bending and H-C stretch modes and 4 quanta in the C-N stretch mode. Due to its nature, HNC is more difficult to produce in the lab thus data is less extensive than for HCN, up to 2 quanta of bend, 4 quanta of H-N stretch and 2 quanta on C-N stretch. This laboratory data is presented in varying formats, from line frequencies to fits of polynomial expansions in angular momentum, such as eq. \ref{eq:rotexp}. Using the electronic data base of laboratory line frequencies from \cite{Maki00,Maki01,Northrup} we have computed and compiled a list of HCN/HNC energy levels relative to the zero point energy of HCN. For any given band some or many line frequencies remain unmeasured, there are therefore missing energy levels in many of the vibrational states. These missing energy levels can be interpolated by means of the rotational expansion for a linear molecule eq. \ref{eq:rotexp}.
For a given vibrational excitation, $l$, and parity we least squares fit eq. \ref{eq:rotexp} to our laboratory determined energy levels, to obtain  $E(v)$, $B_v$, $D_v$, and $H_v$. These constants are then used to interpolate any missing energy levels.

In the high temperatures of a stellar atmosphere the HCN molecule can be excited to a very high angular momentum. HPT found that at 3000~K states up to $J=60$ contribute up to 93\%  of the rotationally converged partition function. The maximum rotational excitation in our laboratory determined linelist is $J=55$, the minimum is $J=20$, to match HPT we must extrapolate to $J=60$. As the rotational expansion (eq. \ref{eq:rotexp}) is divergant, it is not physically realistic and is therfore unsuitable for extrapolation to high $J$. In order to extrapolate to $J=60$, we correct the \ai\  energy levels of HPT by adding a constant for a given vibrational state, $l$, and parity. This constant is given by: $C=E_{lab}(J_{max})-E_{ai}(J_{max})$, where $E_{lab}(J_{max})$ is the energy of the state in the laboratory energy level list with the highest angular momentum ($J_{max}$), and $E_{ai}(J_{max})$, is the energy of the \ai\  energy level with $J=J_{max}$.

The interpolated and extrapolated list of laboratory determined energy levels is referred to as the \leen. It contains 5200 HCN/HNC energy levels up to a maximum of 6 Quanta in bend, 3 quanta of H-C stretch and 3 quanta of C-N stretch and to a maximum angular momentum ($J$) of 60.
We have incorporated the \lee\  energy levels, the 20~000 assignments in to the original \ai\  energy level list of HPT. A sample of this new energy level list is given in table \ref{tab:energies}, the full list can be downloaded from our website (http://www.tampa.phys.ucl.ac.uk/ftp/astrodata ) or the CDS archive (http://cdsweb.u-strasbg.fr/cgi-bin/qcat?/MNRAS/).

The column labelled index in table \ref{tab:energies} is the index number given to the energy level by HPT, $J$ and P are the exact quantum numbers of angular momentum and parity, n is the number of the energy level in the $J$-P symmetry block, E$_{ai}$ is the value of the \ai\ energy level, iso labels the state as either HCN (iso=0) or HNC (iso=1), $v_1$, $v_2$, $l$, and $v_3$ are the approximate quantum numbers, E$_{\rm lab}$ is the \lee\  energy, label is a single character label which is either 'e' for a laboratory determined energy level, 'c' for an interpolated energy level or 't' for a corrected \ai\  energy level. The error on the energy level takes three forms, for a lab determined energy level this is the compound error of the line frequency measurements used to derive the energy level. For an interpolated energy level this is the standard deviation on the fit of eq. \ref{eq:rotexp}. For a corrected \ai\  energy level the error is the difference between the energy predicted by the fit of eq. \ref{eq:rotexp}, to lab and empirical energy levels, and the corrected '\ai' energy. The errors of the lab determined energy levels are typically of the order of $10^{-4}$ \cm, the errors on the interpolated energy levels are typically of the order of $10^{-2}$ \cm, and of the order of 1 \cm\  on the corrected \ai\  energy levels.

\begin{table*}
 \centering
 \begin{minipage}{140mm}
\caption{A sample from the \leen, which is available in full, in electronic form, from either the CDS archive (http://cdsweb.u-strasbg.fr/cgi-bin/qcat?/MNRAS/) or from our website (http://www.tampa.phys.ucl.ac.uk/ftp/astrodata).}
\begin{tabular}{rrrrrrrrrrrrrr}
\hline

Index & $J$ & P & n & E$_{\rm ai}$ (\cm) & iso & $v_1$ & $v_2$ & $l$ & $v_3$ & E$_{\rm lab}$ (\cm) & error (\cm) & label \\ \hline

  1772  & 3 & 1 & 22 & 4704.012753 & 0 & 1 & 2 & 0 & 0 & 4702.019193 &  3.00E-04 & e \\
  1773  & 3 & 1 & 23 & 4719.910847 & 0 & 1 & 2 & 2 & 0 & 4716.909245 &  4.58E-04 & e \\
  1774  & 3 & 1 & 24 & 4880.708077 & 0 & 0 & 7 & 1 & 0 &             &           &   \\
  1775  & 3 & 1 & 25 & 4909.459579 & 0 & 0 & 1 & 1 & 2 & 4895.763446 &  2.00E-04 & e \\
  1776  & 3 & 1 & 26 & 4909.583711 & 0 & 0 & 4 & 0 & 1 & 4906.162828 &  2.04E-01 & c \\
  1777  & 3 & 1 & 27 & 4910.798531 & 0 & 0 & 7 & 3 & 0 &             &           &   \\ 
  1778  & 3 & 1 & 28 & 4925.183847 & 0 & 0 & 4 & 2 & 1 & 4920.596764 &  8.51E-02 & c \\
  1779  & 3 & 1 & 29 & 5203.782210 & 1 & 0 & 0 & 0 & 0 & 5203.779963 &  5.00E-07 & e \\
\end{tabular}
\label{tab:energies}
\end{minipage}
\end{table*}

\section{The new linelist}
\label{sec:linelist}

\begin{figure*}
\includegraphics[angle=-90,width=84mm]{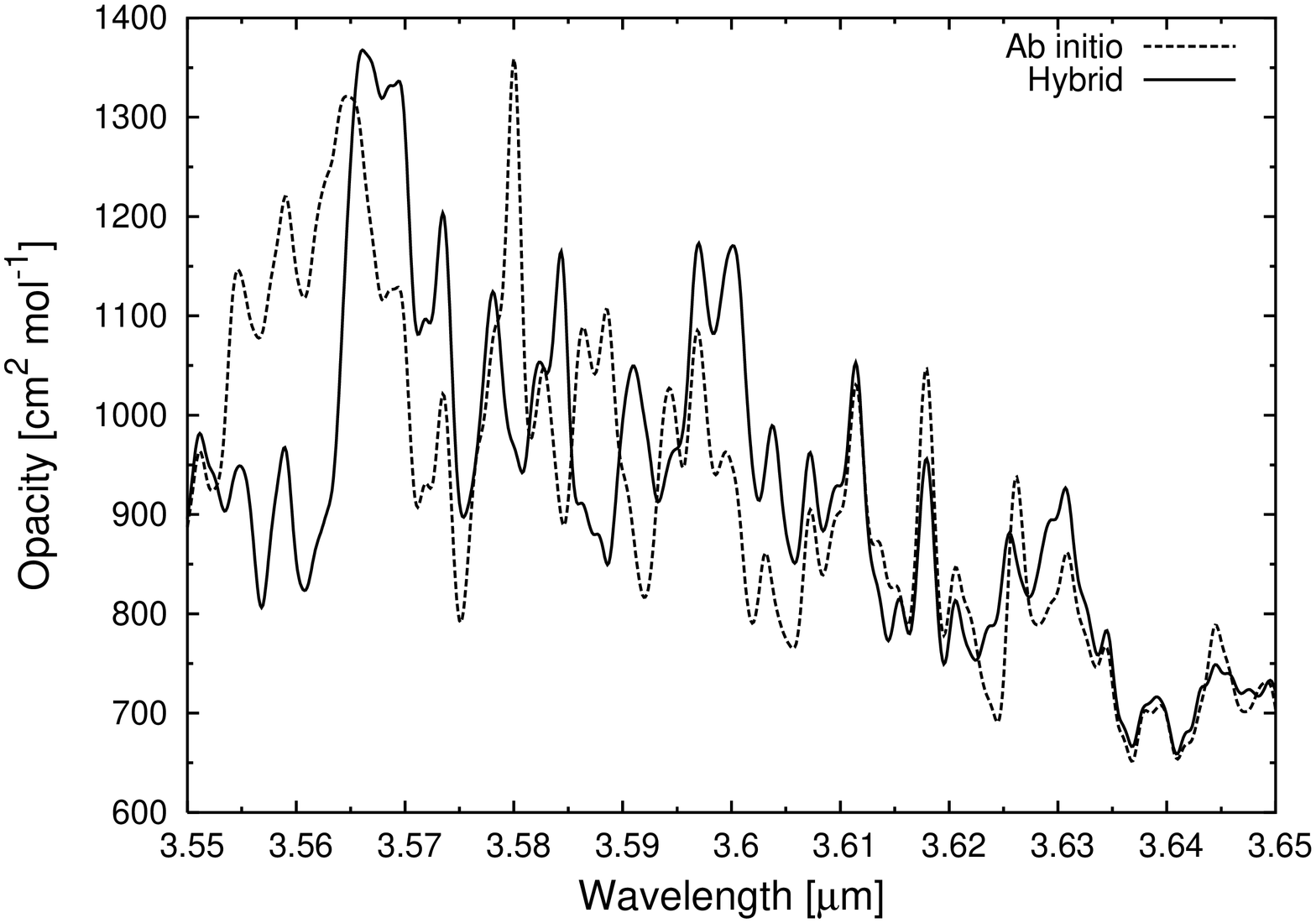}
\caption{The opacity function of HCN/HNC at 2800 K, the lines have been broadened by a Gaussian with half width at half maximum of 0.003 \mim.}
\label{fig:opa3.6}
\end{figure*}

\begin{figure*}
\includegraphics[angle=-90,width=84mm]{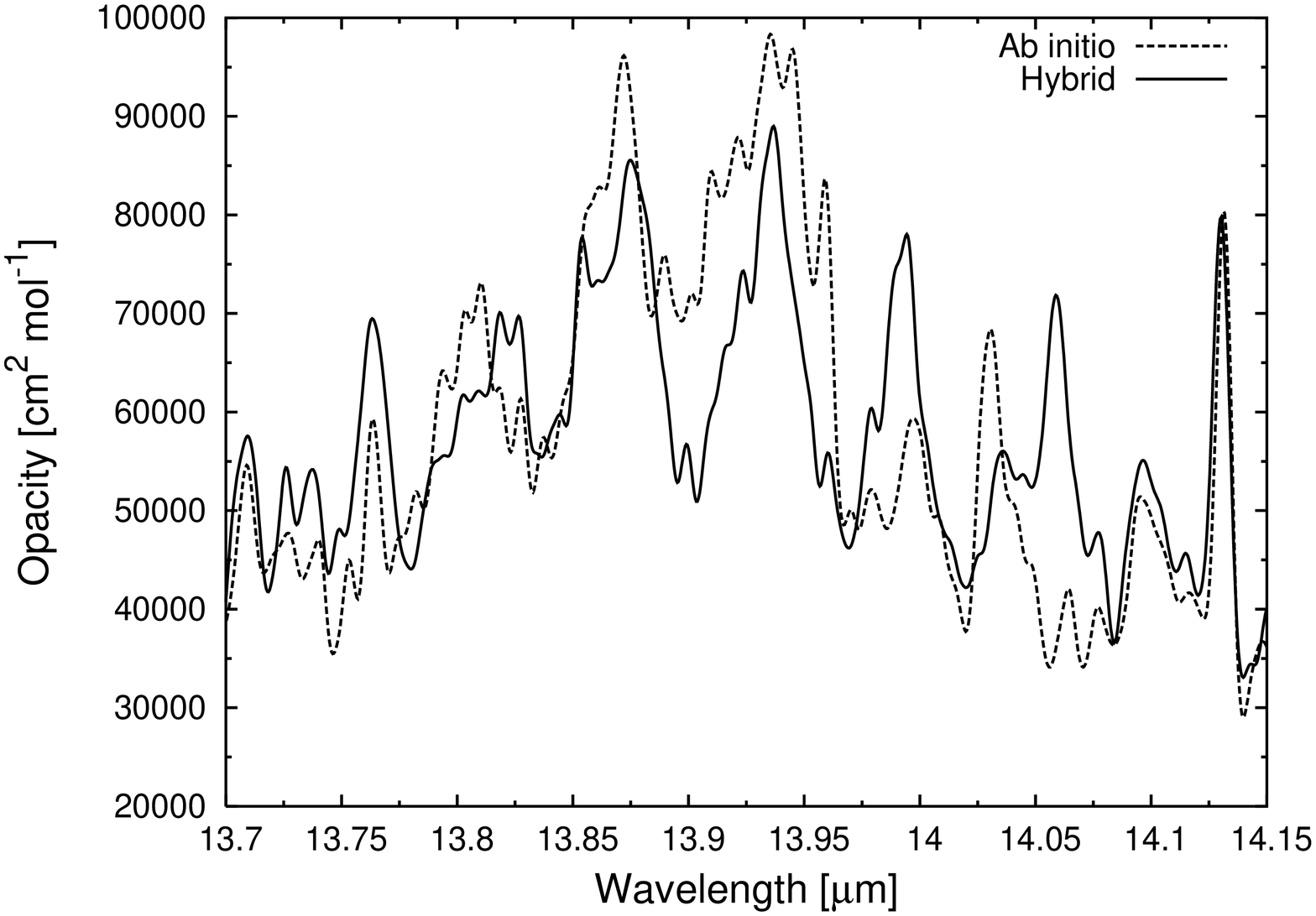}
\caption{The opacity function of HCN/HNC at 2800 K, the lines have been broadened by a Gaussian with half width at half maximum of 0.003 \mim.}
\label{fig:opa14}
\end{figure*}

The \lee\  and \ai\  energy levels, together with the Einstein A coefficients from the  HPT linelist, allow the generation of a new hybrid linelist. For the majority of lines, the line frequencies are computed from the original \ai\   energy levels. However, if \lee\  energy levels have been determined for both the upper and lower states, we can use the \lee\  energy levels to compute a more accurate line frequency. The resulting linelist significantly improves the accuracy of the most intense transitions between the low lying energy levels. This improvement in the linelist is evident in figures \ref{fig:opa3.6} and \ref{fig:opa14}, which show opacity functions calculated with both the \ai\  and hybrid linelists. The predominant features in these opacity functions are the Q branches of the ($\Delta v_2 = 1 + \Delta v_3 = 1$) and ($\Delta v_2=1$) bands, respectively. As the Q branches have a large concentration of lines over a small frequency range, it is with these Q branches that the improvements within the synthetic spectrum are most evident. 

To maintain accuracy and consistency when computing a line frequency it is important to use either \ai\  or \lee\  energy levels for both the upper and lower states.
The error on a given \ai\  energy level in general increases with an increasing value of a quantum number. For example the deviation of the (0,0$^0$,1) $J=0$ \ai\  energy level from laboratory determination is +3.7 \cm, for the (0,0$^0$,2) and (0,0$^0$,3) states the deviation is +8.4 and +13.8 \cm. Thus when calculating frequencies between excited states using the \ai\  energy levels there is, more often than not, a partial cancellation of error. This is not the case if a combination of \lee\  and \ai\  energy levels are used to calculate frequency, which results in a larger error on frequency.

To reduce the required storage space and increase the speed of computations using our linelist, we have truncated the hybrid linelist at a minimum intensity of $3\times10^{-28}$ cm~molecule$^{-1}$ at 3000 K. This results in a linelist of 34.4 million lines less than 10\% of the 393 million lines in the origonal linelist. These 34.4 million lines account for more than 99.99\%\  of the opacity of the full linelist at a temperature of 3000 K. A sample of the linelist is given in table \ref{tab:astrolist}, here $\nu$ is frequency, E\dpr\  is lower state energy, A$_{if}$ is the Einstein A coefficient, $J$\dpr\  and  $J$\pr\  are lower state and upper state angular momentum quntum numbers, p is parity where 1 is even and 0 odd, n is the number of the level in the $J$-parity block, index is the unique label of the energy level, iso labels a HCN state if 0 or a HNC state if 1, $v_1$, $v_2$, $l$, $v_3$ are the approximate quantum numbers. Where the approximate quantum numbers have not been assigned a value of --2 is given.
The full version of the linelist is available from either the CDS archive http://cdsweb.u-strasbg.fr/cgi-bin/qcat?/MNRAS/ or from our website http://www.tampa.phys.ucl.ac.uk/ftp/astrodata. The file size is 3.6~Gb uncompressed and 821~Mb compressed using bzip2. The linelist is sorted is ascending frequency order. 

The untruncated linelist can be obtained by downloading the original HPT Einstein A coefficients, the new assigned energy level list and running the supplied FORTRAN utility program dpsort-v2.0.f90. The Einstein A coefficient file from HPT is sorted using \ai\  frequencies, if lab determined frequencies are used the linelist will no longer be in full frequency order.

The integrated absorption intensity of the lines can be computed from the Einstein A coefficient using
\begin{equation}
I=\frac{C(2J^{\prime}+1)}{Q_{vr}\nu^2}\exp{\left(\frac{-E^{\prime\prime}}{kT}\right)}\left[1-\exp{\left(\frac{-\nu}{kT}\right)}\right]A_{if}
\end{equation}
where $I$ is integrated intensity, $Q_{vr}$ is the ro-vibrational partition function, $J^{\prime}$ is the upper state rotational quantum number, $\nu$ is the transition wavenumber, $E^{\prime\prime}$ is the lower state energy, C is a constant and $A_{if}$ is the Einstein A coefficient. To return $I$ in cm per molecule, with $A_{if}$ in $\rm s^{-1}$ and $\nu$ in \cm\, $C$ has the value of $(8\pi c)^{-1} = 1.3271\times10^{-12}$ s~cm$^{-1}$. The authors recommend the use of the \citet*{Bob} rotationally converged HCN/HNC partition function. Dimensionless oscillator strengths are often used by those involved in the modelling of stellar atmospheres. Weighted oscillator strengths ($gf$) can be computed from the Einstein A coefficients using:
\begin{equation}
gf=\frac{K(2J^{\prime}+1)}{\nu^2}A_{if}
\end{equation}
where $K$ is a constant which for cgs units ($\nu$ in \cm\  and $A_{if}$ in s$^{-1}$) has the value
\begin{equation}
K=\frac{m_e c}{8\pi^2e^2}=1.499197\ \ {\rm s\ cm}^{-2}
\end{equation}
and for SI units ($\nu$ in m$^{-1}$ and $A_{if}$ in s$^{-1}$) the value
\begin{equation}
K=\frac{m_e\epsilon_0c}{2\pi e^2}=14991.97\ \ {\rm s\ m}^{-2}.
\end{equation}

\begin{table*}
 \centering
 \begin{minipage}{180mm}
\vbox to220mm{\vfil Landscape table to go here.}
\caption{}

\label{tab:astrolist}
\end{minipage}
\end{table*}

\section{Computation of model atmospheres and synthetic spectra}

We computed the model atmospheres of C-giants with the SAM12 program 
\citet{Pavlenko2002a,Pavlenko2003} 
which is a modification of ATLAS12 \citep{Kurucz1999}. SAM12 is a plane parallel code
which makes use of the LTE approximation, hydrostatic equilibrium and the conservation 
of flux (there are no sources or sinks of energy within the atmosphere). The 
photospheres of C-giants lie on the upper boundary of their
convective envelopes, thus convection plays an important role in 
these atmospheres. 
We use the mixing length theory of 1D convection modified by
\citet{Kurucz1999} in ATLAS12, and adopt a value of the mixing length parameter of
\lH =1.6.

SAM12 uses the standard set of continuum opacities included in ATLAS12, 
with the addition of continuous opacities from 
C$^-$ \citep{Myerscough1966},
H$_{2}^-$ \citep{Doyle1968}, and bound-free absorption of C, N, O 
from TOPBASE \citep{Seaton1992} with cross-sections 
from \citet{Pavlenko2003c}. These 
bound-free opacity tables have been made available on the Web
\citep{Pavlenko2003d}. Atomic and molecular lines opacities were accounted for
by using the opacity sampling approach \citep{Sneden1976}. 
The data for these lines was obtained from several sources.
HCN/HNC lines taken from this work, atomic lines from the
 VALD database \citep{Kupka1999}, CN, C$_2$, SiH, MgH, CH 
from CDROM 18 of \citet{Kurucz1993}, CO from \citet{Goorvitch1994}, and
CS from \citet{Chandra}.
The following molecular electronic bands 
were accounted for:
CaO (C$^1\Sigma$ - X$^1\Sigma$),  CS(A$^1\Sigma$ - X$^1\Sigma$),
SO (A$^3\Pi$ -  X$^3\Sigma$), SiO (E$^1\Sigma$-X$^1\Sigma$),
SiO(A$^1\Pi$ -  X$^1\Sigma^+$, NO (C$_2 \Pi_r$- X$_2\Pi_r$),
NO(B$_2\Pi_r$ -X$_2\Pi_r$),
NO(A$^2\Sigma^+$ - X$_2\Pi_r$),
MgO(B$^1\Sigma^+$ -  X$^1\Sigma^+$),
AlO(C$^2\Pi$ -X$^2\Sigma$),
AlO(B$^2\Sigma^+$ - X$^2\Sigma^+$),
in the framework
of the JOLA approach. For WZ~Cas these electronic bands are only week 
sources of opacity. For all cases we adopted the carbon isotopic ratio
$^{12}$C/$^{13}$C of 5.
The shape of every line was determined using the Voigt function $H(a,v)$.
Damping constants were taken from line databases or computed using
Unsold's approach \citep{Unsold1955}. In model atmosphere computations
we adopt a microturbulent velocity, V$_{\rm t}$, of
3.5 km/s.

\begin{figure*}
\includegraphics[angle=-90,width=84mm]{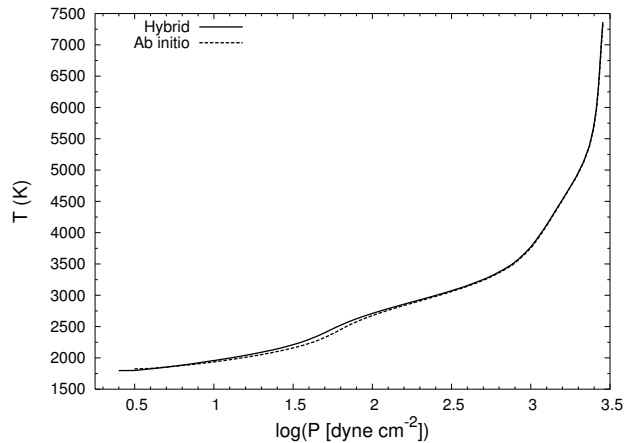}
\caption{Temperature pressure curves of model atmospheres of \teff\  = 2800~K, $\log({\rm N_C}/{\rm N_O}) = -0.003$, $\log(g) = 0.0$, computed with the {\em ab initio} and hybrid linelists.}
\label{fig:atm}
\end{figure*}

Model atmospheres were computed using the 'best fit' effective temperature and C/O ratio, found in our earlier work \citep{paper1}, \teff $=$ 2800~K, $\log({\rm N_C}/{\rm N_O}) = -0.003$, $\log(g) = 0.0$. 
Figure \ref{fig:atm} shows the pressure temperature curves of model atmospheres computed with the \ai\  linelist and hybrid linelist. There is little difference between the atmospheres at depth, but at lower optical depths the model atmosphere computed with the hybrid linelist is slightly hotter at a given pressure.

Synthetic spectra were calculated with the WITA6 program \citep{Pavlenko2000},
using the same approximations and opacities as SAM12, but with a considerably finer 
grid of wavelength points ($\Delta \lambda$ = 0.01 --- 0.02 \AA). We have also computed synthetic spectra with reduced S and Si abundances of [Si/H]=0 and --0.5 dex, to attempt to better fit HCN absorption and reduce the strength of CS absorption, see section \ref{sec:Si}.

\section{Comparison of synthetic spectra with observed stellar spectra.}

The observations of \wzcas\  used in this study were taken as part of the 
Japanese guaranteed 
observing time (REDSTAR1, PI T.Tsuji, e.g., \citet*{Aoki1}). The data and data reduction used in this work is discussed in detail in our earlier paper \citep{paper1}.

\begin{figure*}
\includegraphics[angle=-90,width=84mm]{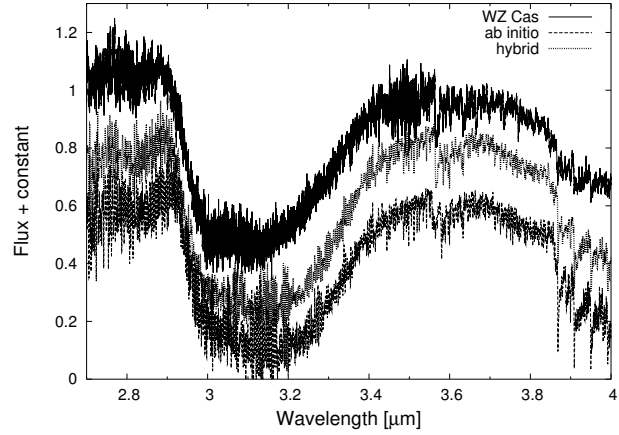}
\caption{Synthetic and observed spectra of \wzcas. The synthetic spectra have been convolved with by a Gaussian with half width at half maximum of 0.003 \mim. The synthetic spectra have constants of 0.2 and 0.4 subtracted from the flux.}
\label{fig:synth-full}
\end{figure*}

\begin{figure*}
\includegraphics[angle=-90,width=84mm]{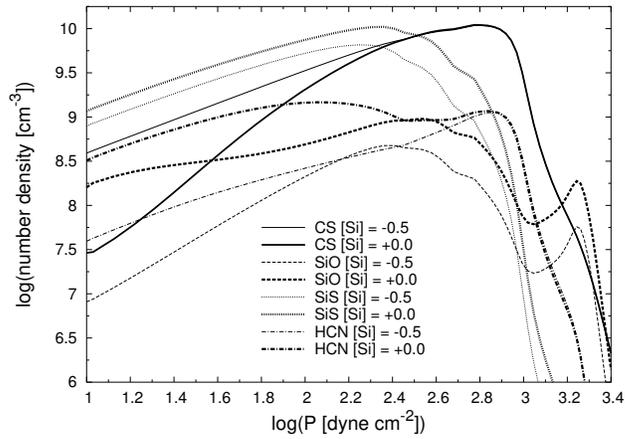}
\caption{Number densities of HCN, CS, SiO and SiS as a function of pressure in the model atmospheres with [Si/H] = 0 and --0.5 dex.}
\label{fig:moleq}
\end{figure*}

Synthetic and observed spectra of \wzcas\  between 2.7 and 4.0 microns are shown in figure \ref{fig:synth-full}. At this resolution there is little difference between the synthetic spectra calculated with the \ai\  and hybrid linelist. 
A synthetic spectrum computed with solar abundances results in
HCN absorption that is stronger than in the observed spectrum WZ Cas.
We find that [Si/H]=-0.5 provides the best fit of HCN features.
The reason for this is that in carbon rich atmospheres silicon and sulphur have a strong secondary affect on chemical equilibrium, after carbon and oxygen \citep{Tsuji}.
Indeed, SiS is analogous to CO and is a very stable molecule, thus atmospheres with a high ratio of Si to S are silicon rich and those with a low ratio are silicon poor. Figure \ref{fig:moleq}, shows the number densities of CS, SiO, SiS and HCN, as a function of pressure in our model atmospheres computed with [Si/H]=0 and --0.5 dex. As the abundance of silicon is reduced from [Si/H]=0 to --0.5 dex, there is less SiS and SiO, freeing S and O to form CO and CS, this reduces the abundance of HCN.

\begin{figure*}
\includegraphics[angle=-90,width=84mm]{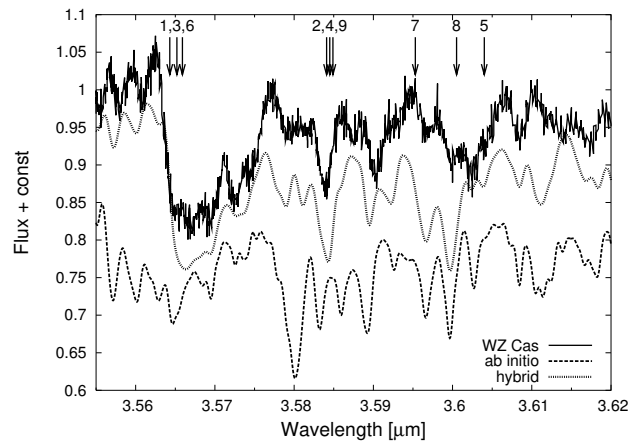}
\caption{Synthetic and observed spectra of \wzcas. The synthetic spectra have been convolved with by a Gaussian with half width at half maximum of 0.003 \mim. The synthetic spectra have constants of 0.1 and 0.2 subtracted from the flux. HCN band centres are labelled with arrows and an index number for the band which is tabulated in table \ref{tab:0111-0000}}
\label{fig:synth-3.6}
\end{figure*}

At higher resolutions the improvements in line positions in the hybrid linelist become more obvious. Figures \ref{fig:synth-3.6} and \ref{fig:synth-3.8} show the synthetic spectra calculated with the \ai\  and hybrid linelists, and the observed spectrum of \wzcas, between 3.55-3.61, and 3.84-3.895 \mum\  respectively. There is a clear and improvement of the fit to observation, across the whole of the 3.55-3.61 \mum\  range. The remaining discrepancies between observed and synthetic spectra over this range are likely to be due either to the spectra of missing chemical species or to transitions between states with $v_2 > 5$, for which laboratory data is unavailable. In figure \ref{fig:synth-3.6}, we have labeled the position of the band centres of several HCN bands. Absorption features can be seen close to the band centres, these are due to the Q-branch ($\Delta J=0$) lines of the band, which are closely spaced and result in strong absorption over a narrow range. 

\begin{table*}
 \centering
 \begin{minipage}{140mm}
\caption{Laboratory band centres for bands with $\Delta v_2=1$,$\Delta v_3=1$.}
\begin{tabular}{cccc}
\hline
index \footnote{This is the index of the band used in figure \ref{fig:synth-3.6}} &
$(v_1^\prime,v_2^\prime,l^\prime,v_3^\prime)$ & 
$(v_1^{\prime\prime},v_2^{\prime\prime},l^{\prime\prime},v_3^{\prime\prime})$ & $\lambda_c$ \footnote{From the laboratory measurements of \citet{Maki96}.}
   \mim \\ \hline

 1 & (0,1,1,1)  & (0,0,0,0) & 3.5643 \\
 2 & (0,2,0,1)  & (0,1,1,0) & 3.5841 \\
 3 & (0,2,2,1)  & (0,1,1,0) & 3.5652 \\
 4 & (0,3,1,1)  & (0,2,0,0) & 3.5845 \\
 5 & (0,3,1,1)  & (0,2,2,0) & 3.6040 \\
 6 & (0,3,3,1)  & (0,2,2,0) & 3.5659 \\
 7 & (0,1,1,2)  & (0,0,0,1) & 3.5953 \\
 8 & (0,4,0,1)  & (0,3,1,0) & 3.6041 \\
 9 & (0,4,2,1)  & (0,3,1,0) & 3.5849 \\
\end{tabular}

\label{tab:0111-0000}
\end{minipage}
\end{table*}

\subsection{The CS $\Delta v=2$ band heads and the abundance of Sulphur and Silicon.}
\label{sec:Si}

\begin{figure*}
\includegraphics[angle=-90,width=84mm]{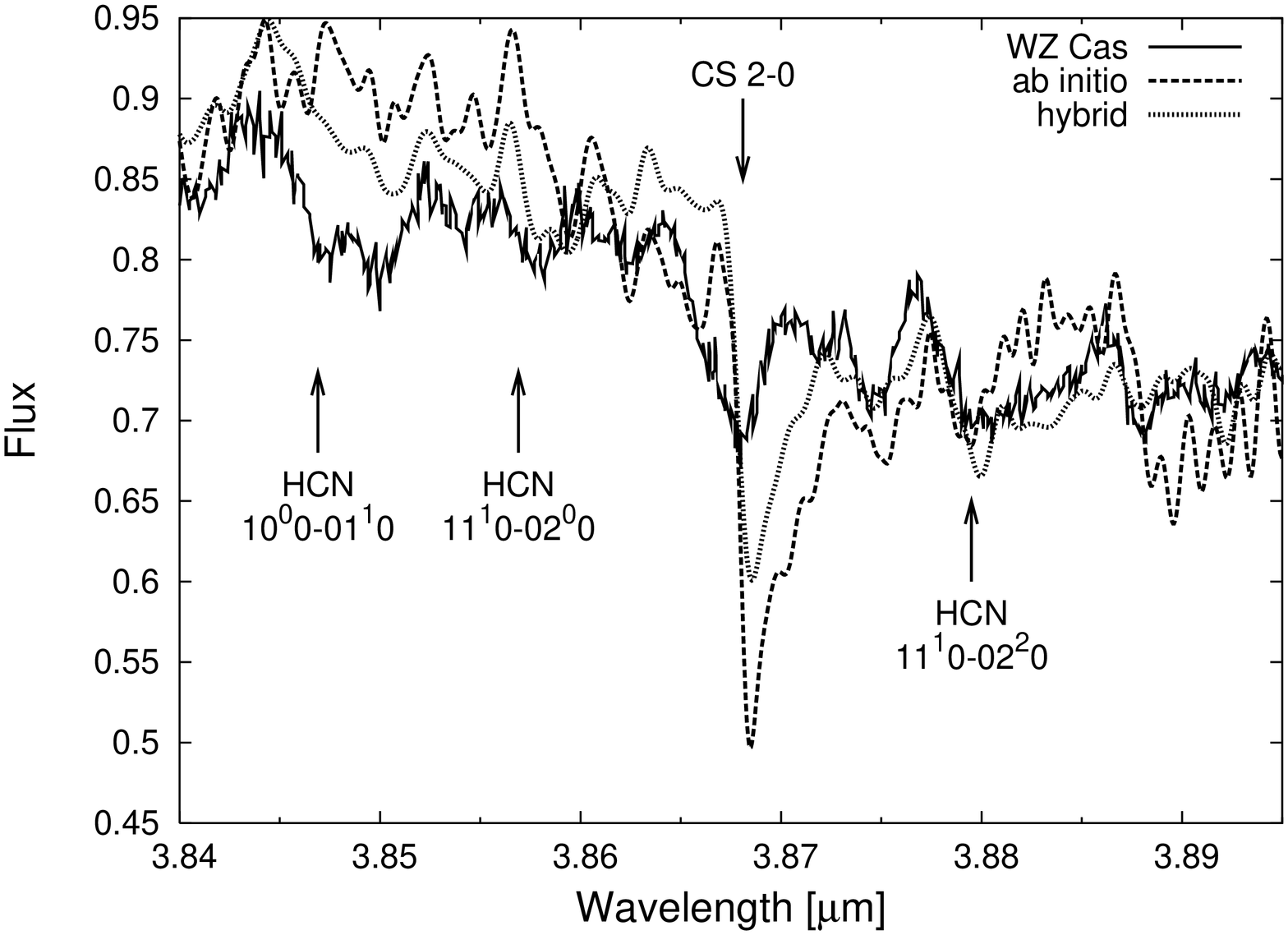}
\caption{Synthetic and observed spectra of \wzcas. The synthetic spectra have been convolved with by a Gaussian with half width at half maximum of 0.003 \mim.}
\label{fig:synth-3.8}
\end{figure*}

In the 3.84-3.895 \mum\  region shown in figure \ref{fig:synth-3.8}, we have identified the Q-branches of the 3 lowest lying HCN states. The reproduction of the shapes of these HCN absorption features is significantly improved, by the use of the hybrid linelist. However, we over-predict the strength of the CS $v=0\rightarrow2$ band head at about 3.868 \mum. \cite{Aoki1} had the same problem, they suggested that the poor fit was a result of emission by CS from a circumstellar shell. We feel that this is an unlikely scenario. Absorption from both the fundamental and the hot bands is clear in our synthetic spectra, and also appear, all be it more weakly, in the observed spectrum. In contrast, emission from CS at temperature significantly lower than that of the stellar photosphere, will, be predominantly from the fundamental and the low lying hot bands. Thus the signature of atmospheric absorption with emission from a cool circumstellar shell would be a weak fundamental band and strong hot bands, this is not seen in the observed spectra.

\begin{figure*}
\includegraphics[angle=-90,width=84mm]{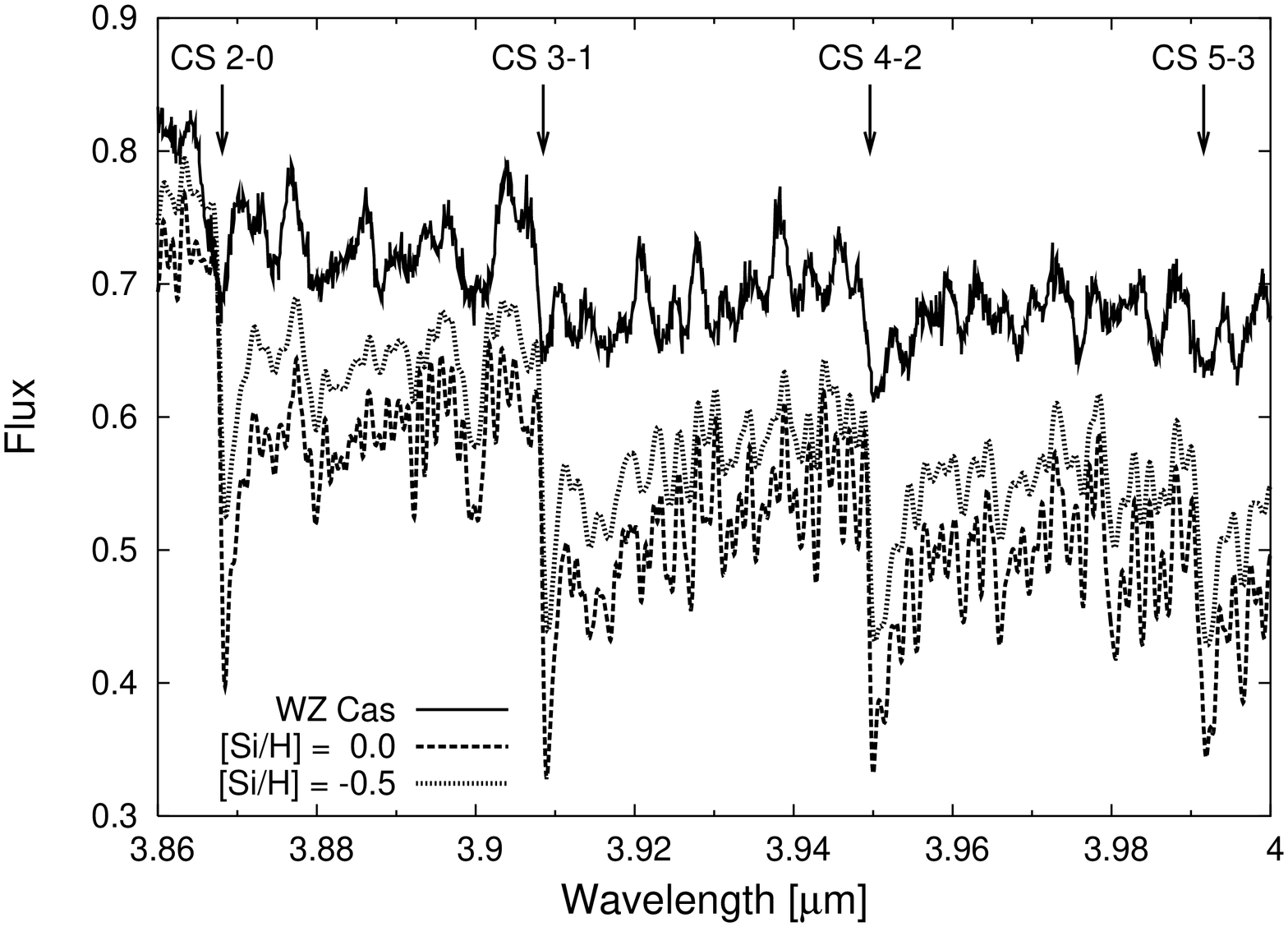}
\caption{Synthetic and observed spectra of \wzcas. The synthetic spectra have been calculated with different Si abundances and are convolved with by a Gaussian with half width at half maximum of 0.003 \mim. The synthetic spectra have constants of 0.075 and 0.1 subtracted from the flux.}
\label{fig:synth-3.8-Si}
\end{figure*}

As an alternative to circumstellar shell emission wetested alternative metal mixes. The problem, however, cannot be solved by reducing the sulphur abundance, because not only are the strength of the CS lines reduced, but chemical equilibrium is altered resulting in a greater abundance of HCN, so that the HCN features from 2.8-4 \mum\  were poorly reproduced. Similarly increasing the abundance of silicon to [Si/H]=0 dex weakens slightly the CS features, but increases the stength of HCN absorption. The net effect is little improvement in the reproduction of absorption over the 3.8-4 micron region, see figure \ref{fig:synth-3.8-Si}, but a worse reproduction of the HCN absorption accross the wider spectral range.

We use the CS linelist of \citet{Chandra} in our computations, we do not believe that errors in this linelist contribute to the over prediction of the strength of CS features in our synthetic spectra. This CS linelist was calculated by using Dunham coefficients from \citet{Winkel}, and the dipole moment surface of \citet{Botschwina}. Within the rigid rotor approximation, the rotational $J=$0$\rightarrow$1 transition dipole is equal to the permanent dipole moment, thus can be used as a rough check for the accuracy of intensity calculations. The $J=$0$\rightarrow$1 transition dipole computed by \citet{Chandra} is 1.889 D, the permanent dipole moment quoted by \citet{Botschwina} is 1.958~D. The vibrational band transition dipoles given by \citet{Botschwina} are 0.1580~D and 0.00911~D for the $v=0\rightarrow1$ and $v=0\rightarrow2$ transitions respectively. The $v=0\rightarrow1$ and $v=0\rightarrow2$ transitions given by \citet{Chandra} are 0.1569~D and 0.00788~D.
Although there are small differences, this is not large enough to account for the difference between observation and synthetic spectra. A final possibility is that there is an opacity source present in the stellar atmosphere which we  have not accounted for in our models. This could be from a species such as C$_2$H$_2$, which is known to exist in some C-star atmospheres, but is not included in our calculations, as no C$_2$H$_2$ linelist is available. It is possible that increased opacity from such species could affect both the structure of model atmospheres and/or change the synthetic spectra.

\section{Conclusion.}
\label{sec:conc}

We have assigned approximate quantum numbers to 20~000 of the {\em ab initio} energy levels found in the HCN/HNC linelist of \citet{linepaper}. Existing laboratory measurements of HCN and HNC line frequencies are used to accurately determine 5200 energy levels. Both the energy level assignments and the \lee\  energy levels have been included along with the \ai\  energy levels of \citet{linepaper} in a new energy level list. A combination of the \lee\  and \ai\  energy levels are used in conjunction with the \citet{linepaper} Einstein A coefficients to produce an improved linelist. We make the new energy level file which includes assignments, laboratory determined energy levels and the resulting improved HCN and HNC linelist publicly available.

The new linelist has been incorporated into our computations of C-rich model stellar atmospheres and synthetic spectra. The new synthetic spectra are a clear improvement over our earlier synthetic spectra \citep{paper1}, and show better agreement with the observed spectrum of WZ~Cas. The improvement is clearest in the range 3.56-3.62 \mum. We further improve the fit to HCN features by reducing the Si abundance to [Si/H]=--0.5 dex, however the CS absorption feature at 3.8-4.0 microns is still poorly reproduced.

\section*{Acknowledgements}

GJH thanks the UK Particle Physics and Astronomy Research Council (PPARC), for post-doctoral funding. The work YP is partially supported by the Leverhume Trust. The manipulation and analysis of the HCN/HNC linelist was carried out on the Enigma computer facility of the HiPerSPACE computing centre at UCL which is part funded by PPARC.

\begin{figure*}
\includegraphics[angle=-180,width=200mm]{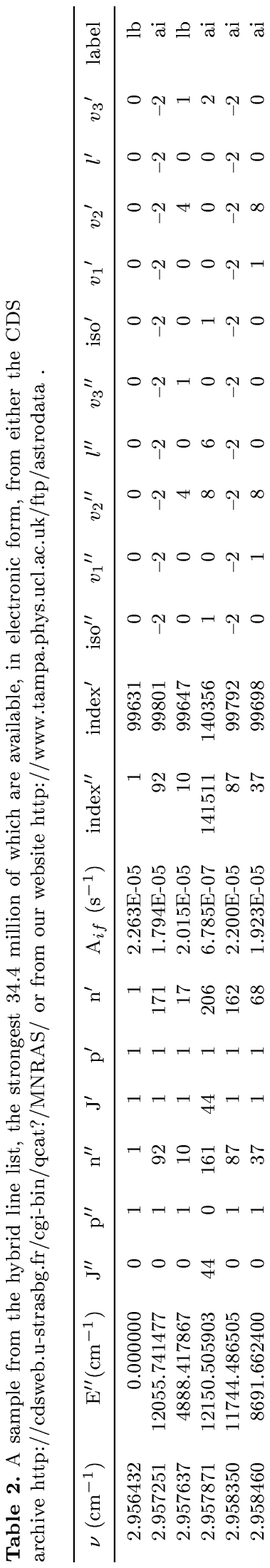}
\caption{Demo of lanscape table.}
\end{figure*}

\end{document}